\documentclass[a4paper,11pt]{article}
\usepackage{pos}
\usepackage{comment}

\usepackage[dvipsnames]{xcolor}

\usepackage[normalem]{ulem}

\newcommand{\pjt}{p_{jT}}
\newcommand{\jt}{j_T}
\newcommand{\etaj}{\eta_j}
\newcommand{\sqs}{\sqrt{s}}
\newcommand{\pperp}{p_{\perp h}}
\newcommand{\hone}{h_1}

\newcommand{\WW}{\mathrm{WW}}
\newcommand{\LO}{\mathrm{LO}}

\title{Collins effect in pion-in-jet production in polarized $pp$ and $ep$ collisions}
\ShortTitle{Collins effect in pion-in-jet production in polarized $pp$ and $ep$ collisions}

\author*[a,b]{Carlo Flore}
\author[a,b]{Umberto D'Alesio}
\author[c]{Marco Zaccheddu}

\affiliation[a]{Dipartimento di Fisica, Universit\`a di Cagliari,\\
  Cittadella Universitaria, I-09042 Monserrato (CA), Italy}

\affiliation[b]{INFN, Sezione di Cagliari,\\
  Cittadella Universitaria, I-09042 Monserrato (CA), Italy}

\affiliation[c]{Theory Center, Jefferson Lab,\\
  12000 Jefferson Avenue, Newport News, Virginia 23606, USA}

\emailAdd{carlo.flore@unica.it}
\emailAdd{umberto.dalesio@ca.infn.it}
\emailAdd{zacch@jlab.org}

\abstract{We study Collins azimuthal asymmetries for pion-in-jet production in polarized proton-proton and lepton-proton collisions. We adopt a hybrid transverse momentum dependent  approach, with a collinear configuration for the initial state, and employ the transversity and Collins fragmentation functions extracted from semi-inclusive deep inelastic scattering and $e^+e^-$ annihilation data. After recalling the good description of the STAR data in $pp$ collisions, which supports the universality of the Collins function, we present predictions for Electron-Ion Collider kinematics, both at leading order  and by including the quasireal photon exchange in the Weizs\"acker-Williams approximation. This contribution is sizable but does not spoil the dominance of quark-initiated channels. This implies that $\ell p$ processes allow for a clearer access to the transversity distribution, including its sea-quark component.}

\FullConference{The 33rd International Workshop on Deep Inelastic Scattering and Related Subjects (DIS2026)\\
4 -- 8 May 2026\\
Bologna, Italy\\}

\begin{document}
\renewcommand{\hookAfterAbstract}{%
\par\bigskip
%\textsc{ArXiv ePrint}:
%\href{https://arxiv.org/abs/1234.5678}{1234.5678}
\textsc{Report n.~JLAB-THY-26-4821}
}
\maketitle

\section{Introduction}
\label{sec:intro}
Mapping the three-dimensional structure of hadrons is a central goal in understanding how hadron properties emerge from the dynamics of confined quarks and gluons. Azimuthal and transverse-spin asymmetries are a powerful tool in this regard, giving access to spin--transverse-momentum correlations both in the nucleon and in the hadronization mechanism. The transverse momentum dependent (TMD) approach~\cite{Collins:1981uk,Collins:1984kg} provides the consolidated framework to describe these observables, one of its key ingredients being the presence of two ordered energy scales, as in the case of semi-inclusive deep inelastic scattering (SIDIS), Drell-Yan and back-to-back hadron-pair production in $e^+e^-$ collisions.

Among the various TMD distributions, a genuine TMD fragmentation function (FF) of central interest here is the Collins function~\cite{Collins:1992kk}, which describes the fragmentation of a transversely polarized quark into an unpolarized hadron and correlates the quark transverse polarization with the transverse momentum of the produced hadron. Together with the transversity distribution $\hone^q$, it is extracted from global fits of SIDIS and $e^+e^-$ azimuthal asymmetries~\cite{Anselmino:2007fs,DAlesio:2020vtw,Boglione:2024dal}. A crucial property of this TMD FF is its predicted universality when moving from SIDIS and $e^+e^-$ processes~\cite{Metz:2002iz,Collins:2004nx}, in contrast to the sign change of the Sivers function between SIDIS and Drell-Yan.

The production of a light hadron inside a jet offers a further, independent testing ground for both TMD factorization and Collins-function universality~\cite{Yuan:2007nd,DAlesio:2010sag,Kang:2017btw}. Here the two ordered scales are the large jet transverse momentum $\pjt$ and the small transverse momentum $\jt$ of the hadron with respect to the jet axis, with TMD effects are restricted only in the fragmentation sector. The relevant observable is the Collins azimuthal asymmetry
\begin{equation}
A_{UT}^{\sin(\phi_S-\phi_h^H)}
= \frac{2\int d\phi_S\, d\phi_h^H \,\sin(\phi_S-\phi_h^H)\,
        \big[d\sigma(\phi_S,\phi_h^H) - d\sigma(\phi_S+\pi,\phi_h^H)\big]}
       {\int d\phi_S\, d\phi_h^H \,
        \big[d\sigma(\phi_S,\phi_h^H) + d\sigma(\phi_S+\pi,\phi_h^H)\big]}\,,
\label{eq:asy}
\end{equation}
whose numerator, schematically, is a convolution of the collinear transversity function, the polarized partonic cross section, and the Collins TMD FF, $N[A_{UT}]\sim\sum \hone^a\otimes f_1^b\otimes \Delta\hat\sigma^{ab\to cd}\otimes H_1^{\perp\,c}$. 

In what follows, we first summarize the analysis of the $pp$ case, and then present our latest results for lepton-proton collisions at the EIC. For these studies, we adopted the parametrization of $\hone^q$ and $H_1^{\perp\,q}$ from Ref.~\cite{Boglione:2024dal}. As collinear inputs, the MSHT20nlo proton PDFs~\cite{Bailey:2020ooq} and the DEHSS pion FFs~\cite{deFlorian:2014xna} are taken. The transversity function is evolved with a modified version of {\tt HOPPET}~\cite{Salam:2008qg}, and the hard scale is set to $\mu=\pjt$.

\section{Pion-in-jet production in $pp$ collisions}
\label{sec:pp}
\begin{figure}[t]
\centering
\includegraphics[width=.49\textwidth]{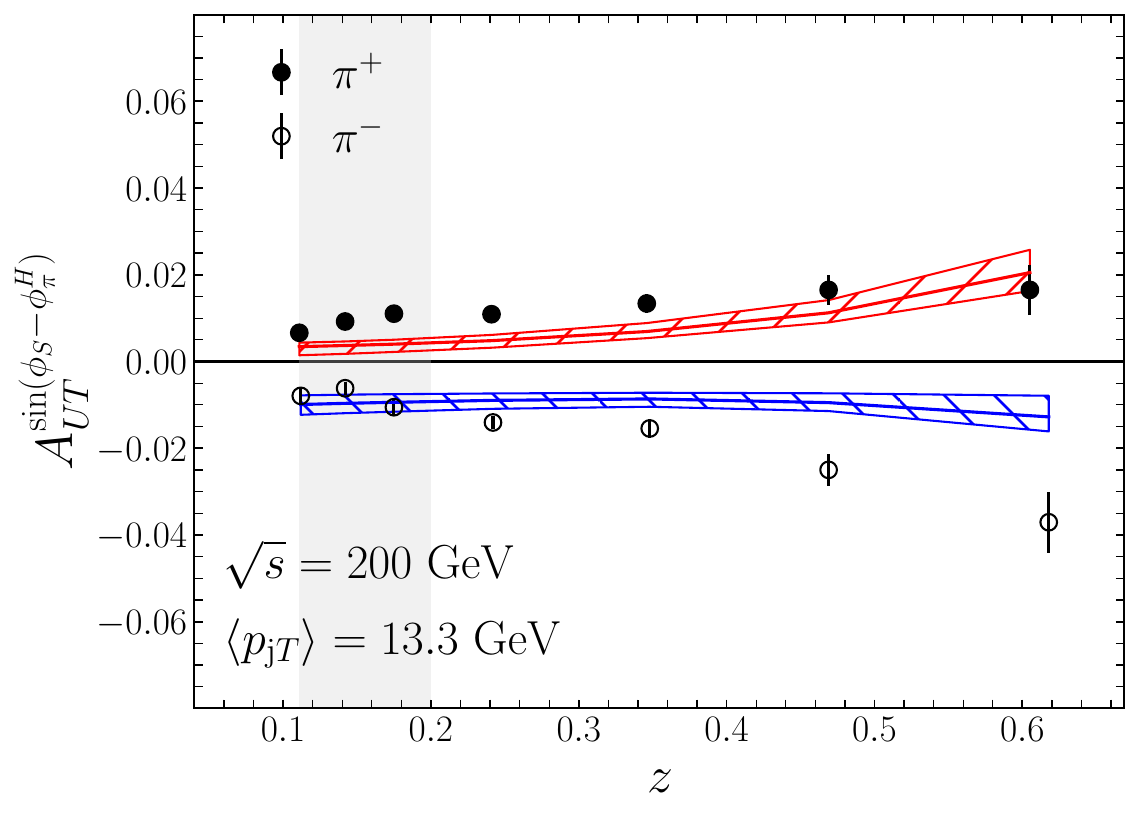}
\includegraphics[width=.49\textwidth]{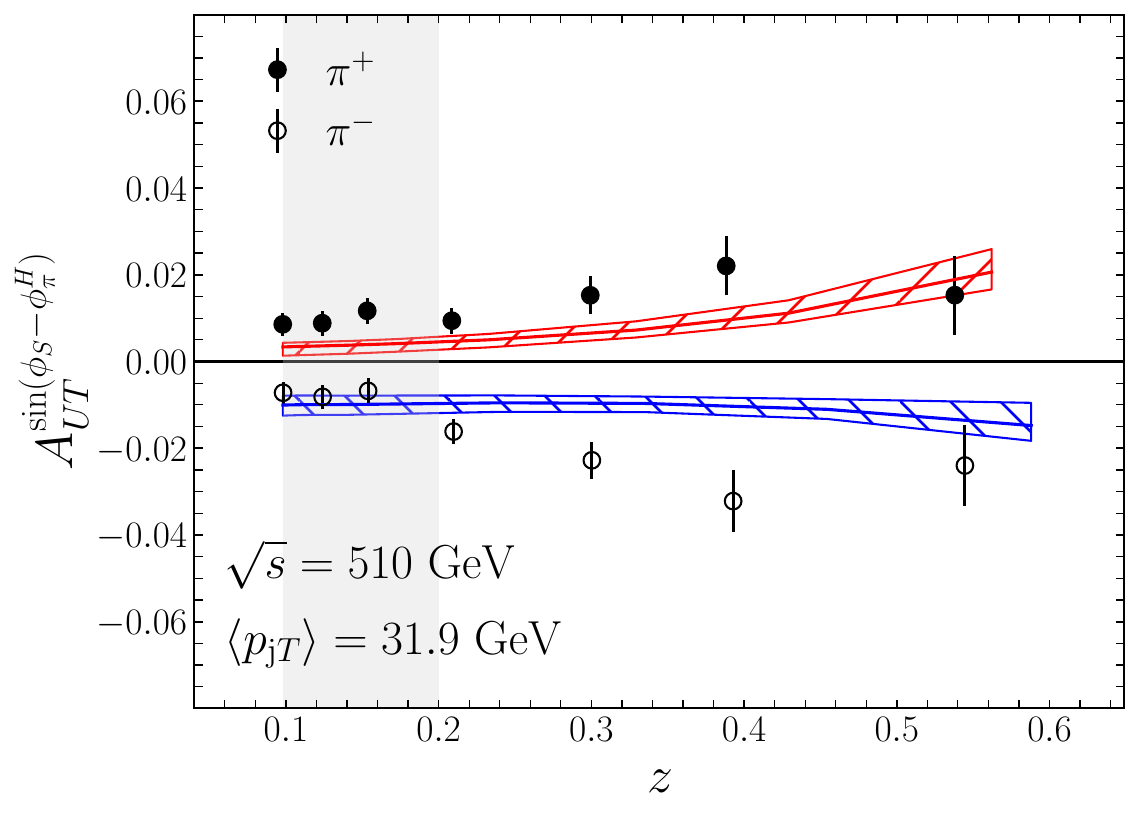}
\caption{Collins asymmetry $A_{UT}^{\sin(\phi_S-\phi_\pi^H)}$ for $p^\uparrow p \to {\rm jet}\,\pi^\pm X$ as a function of $z$, compared to STAR data at $\sqs=200$ (left)~\cite{STAR:2022hqg} and $510$~GeV (right)~\cite{STAR:2025xyp}. Postdictions are obtained using the transversity and Collins functions of Ref.~\cite{Boglione:2024dal}. The shaded grey area marks the region not covered by the SIDIS and $e^+e^-$ data used in the extraction. Figure taken from Ref.~\cite{DAlesio:2025jmr}.}
\label{fig:star}
\end{figure}

In Ref.~\cite{DAlesio:2025jmr} we analyzed the Collins asymmetry for $p^\uparrow p \to {\rm jet}\,\pi\,X$ within a LO TMD approach, and with a collinear QCD configuration for the initial state. %, using the transversity and Collins functions from the updated simultaneous reweighting of SIDIS and $e^+e^-$ data of Ref.~\cite{Boglione:2024dal}. 
As shown in Fig.~\ref{fig:star}, our postdictions based on Ref.~\cite{Boglione:2024dal} provide a good description of the STAR data~\cite{STAR:2022hqg,STAR:2025xyp} at two center-of-mass (c.m.) energies, $\sqs=200$ and $510$~GeV, over the entire $z$ range (and similarly for the $x_T$ and $\jt$ dependencies, cfr.~Figs.~1 and 4 of Ref.~\cite{DAlesio:2025jmr}). The shaded band highlights the small-$z$ region that is not covered by the kinematics of the SIDIS and
$e^+e^-$ data entering the global extraction. % A subsequent reweighting on the STAR asymmetries themselves, performed both in the generalized parton model and in its color-gauge-invariant version, further improves the agreement at both energies.

The main outcomes of this study are the good overall data-theory agreement, % the absence of sizable factorization-breaking effects,
the marginal role of TMD evolution and, overall, a support for the universality of the Collins FF and for TMD factorization in this class of processes. Although the STAR data already indicate that hadron-in-jet production can improve and extend our knowledge of TMDs at large $x$, these findings motivate extending the analysis to the complementary $\ell p$ process.

\section{Pion-in-jet production in $\ell p$ collisions}
\label{sec:eic}
The process $\ell p^\uparrow \to {\rm jet}\,h\,X$, where the final-state lepton not observed, is simpler than and complementary to its $pp$ counterpart~\cite{DAlesio:2026pca}. At LO, $\mathcal{O}(\alpha^2)$, it involves only the partonic channel $q\ell\to q\ell$, with no gluon contribution. Therefore, the denominator of the Collins asymmetry is not affected by gluon-induced channels, that can be sizeable in the $pp$ case, allowing for a cleaner access to the transversity function. %Writing $\hone^q$ from the valence-like parametrization of Ref.~\cite{Boglione:2024dal} and the Collins FF in its favored/unfavored decomposition, 

The LO unpolarized and polarized cross sections read schematically
\begin{align}
2\,d\sigma^{\rm unp}_{\LO} &\sim \sum_q f_{q/p}(x)\otimes d\hat\sigma_{q\ell\to q\ell}\otimes
  D_{h/q}(z,\pperp^2)\,,\\
d\Delta\sigma_{\LO} &\sim \sin(\phi_S-\phi_h^H)\sum_q \hone^{q}(x)\otimes
  d\Delta\hat\sigma_{q\ell\to q\ell}\otimes H_1^{\perp\,q}(z,\pperp^2)\,.
\end{align}
%where $\Delta^N\! D_{h/q^\uparrow}$ is the Collins TMD-FF. .

%\paragraph{Weizs\"acker-Williams contribution.}
Since the cross section is dominated by $Q^2\simeq 0$, we included at $\mathcal{O}(\alpha^2 \alpha_s)$ the quasireal photon exchange in the Weizs\"acker-Williams approximation~\cite{vonWeizsacker:1934nji,Williams:1934ad}, in which the lepton acts as a source of real photons. The corresponding contribution to the cross section is
\begin{equation}
d\sigma^{\WW}(\ell N\to {\rm jet}\,h\,X) = \int dy\; f_{\gamma/\ell}(y)\;
  d\sigma(\gamma N \to {\rm jet}\,h\,X)\,,
\label{eq:WWfact}
\end{equation}
with the photon flux~\cite{Kniehl:1996we}
\begin{equation}
f_{\gamma/\ell}(y) = \frac{\alpha}{2\pi}\!\left[
  \frac{1+(1-y)^2}{y}\,\ln\frac{Q^2_{\max}}{Q^2_{\min}(y)}
  + 2m_\ell^2\,y\left(\frac{1}{Q^2_{\max}}-\frac{1}{Q^2_{\min}(y)}\right)\right],
\label{eq:flux}
\end{equation}
where $Q^2_{\max}=1~\mathrm{GeV}^2$ and $Q^2_{\min}(y)=m_\ell^2 y^2/(1-y)$. 
The inclusion of the WW term opens new channels in the unpolarized cross section: (i) $q\gamma\to qg$, (ii) $q\gamma\to gq$ and (iii) $g\gamma\to q\bar q$. The corresponding partonic cross section are computed with the helicity amplitudes taken from Ref.~\cite{DAlesio:2017nrd}. The numerator of the asymmetry, however, receives a contribution only from $q\gamma\to qg$ scattering, since all other spin transfers vanish and no gluon transversity exists for a spin-$1/2$ target~\cite{DAlesio:2017nrd}.

\paragraph{Predictions for the EIC}

\begin{figure}[t]
\centering
\includegraphics[width=0.90\textwidth]{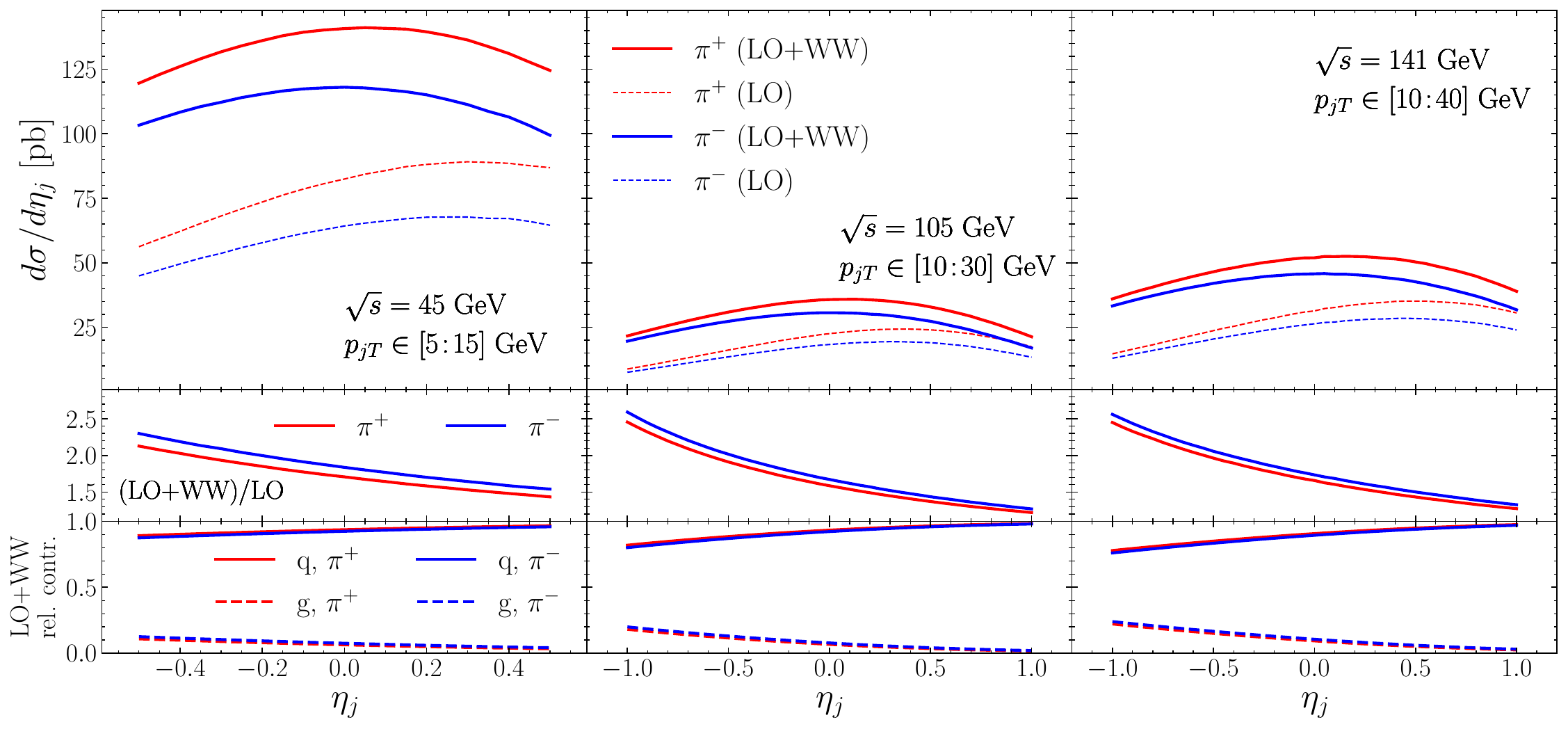}
\caption{Unpolarized cross sections for $\ell p \to {\rm jet}\,\pi^\pm X$ as a function of $\etaj$ at $\sqs=45,\,105,\,141$~GeV, with the corresponding $\pjt$ integration ranges. Upper panels: LO and LO+WW results; middle panels: ratio of the full (LO+WW) result to the LO piece; lower panels: relative contribution of quark- and gluon-induced channels to the total cross section. Figure taken from Ref.~\cite{DAlesio:2026pca}.}
\label{fig:xsec}
\end{figure}

%\paragraph{Role of the WW term.}
We present results for the unpolarized cross section for $\pi^\pm$ production at three different c.m.~energies, $\sqs=45,\,105,\,141$~GeV. The corresponding $\pjt$ integration ranges $[5{:}15]$, $[10{:}30]$ and $[10{:}40]$~GeV are chosen to explore comparable ranges in $x_T=2\pjt/\sqs$. As shown in Fig.~\ref{fig:xsec}, the WW contribution significantly enhances the unpolarized cross section: the ratio of the full (LO+WW) result to the LO one is about $2$ in the backward and $1.5$ in the forward region, with gluon-induced channels playing a role mostly in the backward region. Nonetheless, the quark-induced contribution remains dominant, being $80$--$90\%$ of the cross section. Since only quark-initiated channels enter the numerator, this marginal gluon role in the denominator, at variance with the $pp$ case, is what allows for a clean access to transversity.

\begin{figure}[t]
\centering
\includegraphics[width=0.90\textwidth]{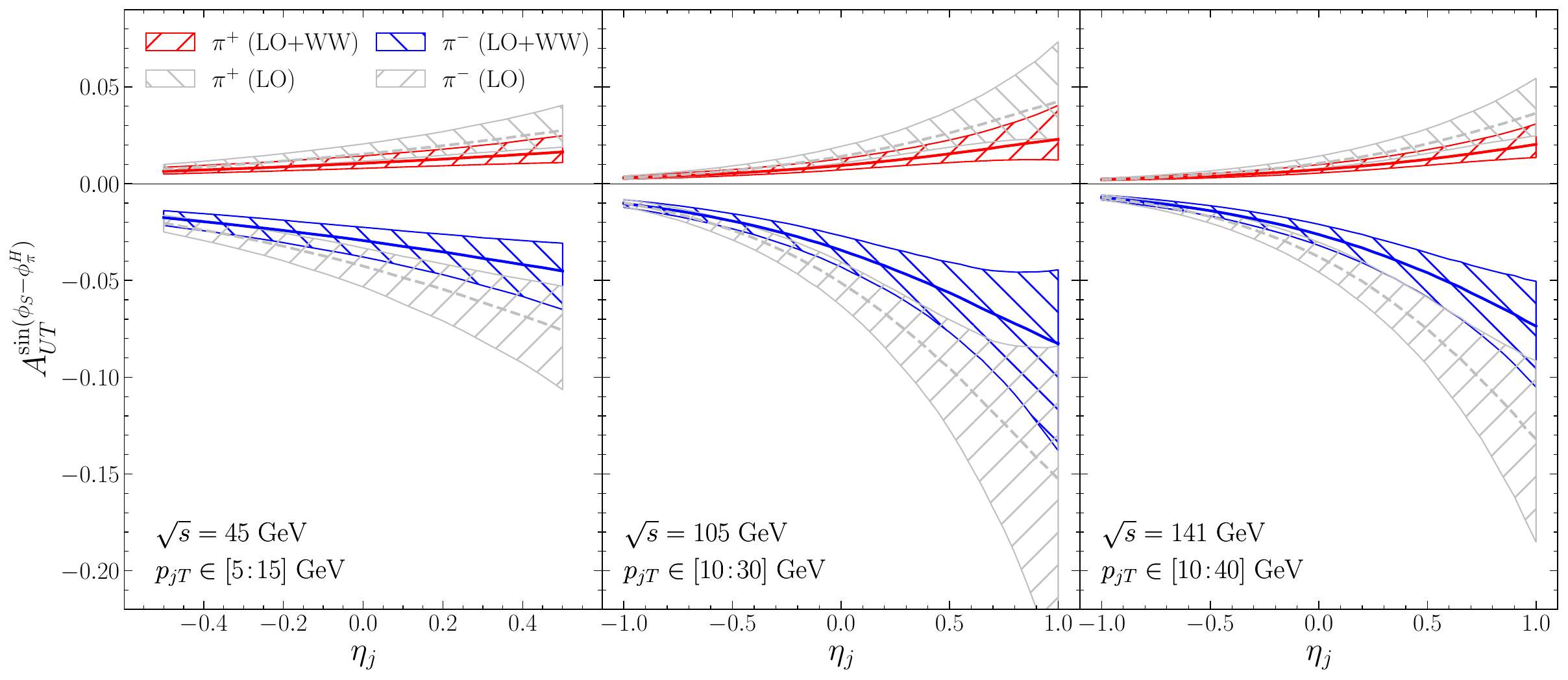}
\caption{Collins asymmetries for $\ell p^\uparrow \to {\rm jet}\,\pi^\pm X$ as a function of $\etaj$: LO (dashed lines, gray bands) and LO+WW (solid lines, colored bands). From left to right: $\sqs=45,\,105,\,141$~GeV, with the corresponding $\pjt$ ranges. Uncertainty bands at $2\sigma$ CL. Figure taken from Ref.~\cite{DAlesio:2026pca}.}
\label{fig:eic_eta}
\end{figure}

%\paragraph{Predictions for the Collins asymmetry at the EIC.}
Fig.~\ref{fig:eic_eta} shows our predictions for the Collins asymmetry as a function of $\etaj$, with median as central value uncertainty at $2\sigma$ confidence level (CL). The asymmetry decreases in size when including the WW contribution, especially at forward rapidity. In this same region, the asymmetry is larger: about $1-2\%$ for $\pi^+$, and $4-8\%$ for $\pi^-$. This follows from the partial suppression of the favored Collins FF once integrated over $z$, and from the increasing role of the up-quark transversity at larger $x$. At large negative pseudorapidities and at larger c.m.~energies, one probes small-$x$ values ($x\sim 10^{-2}$), and our valence-like transversity parametrization yields strongly suppressed asymmetries. This kinematic region is therefore particularly sensitive to the sea-quark component of transversity, which is still largely unconstrained by current data.

\begin{figure}[t]
\centering
\includegraphics[width=0.8\textwidth]{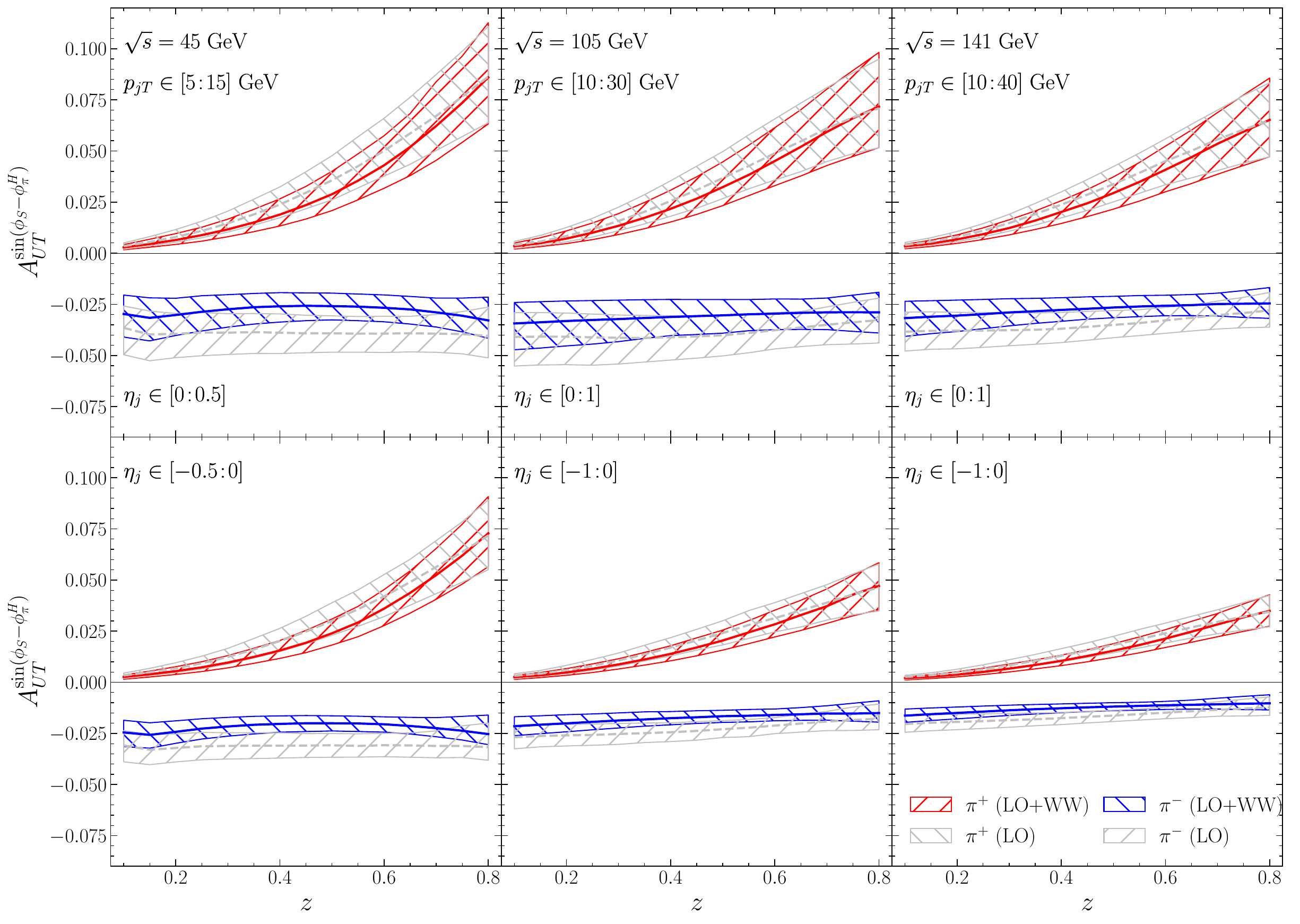}
\caption{Collins asymmetries for $\ell p^\uparrow \to {\rm jet}\,\pi^\pm X$ as a function of $z$: LO (dashed lines, gray bands) and LO+WW (solid lines, colored bands). From left to right: $\sqs=45,\,105,\,141$~GeV. Upper panels: forward rapidities; lower panels: backward rapidities. Uncertainty bands at $2\sigma$ CL. Figure taken from Ref.~\cite{DAlesio:2026pca}.}
\label{fig:eic_z}
\end{figure}

Finally, in Fig.~\ref{fig:eic_z}, we show predictions for the $z$ dependence of the Collins asymmetry. For $\pi^-$ the asymmetry is almost flat, around $2 - 3\%$, whereas for $\pi^+$ it increases with $z$, reaching about $8\%$ in the forward region. This reflects the dominance of the up-quark contribution (large transversity and charge factor) combined with the suppression of the unfavored Collins FF at large $z$: the $\pi^+$ numerator is driven by the up-quark transversity times the favored Collins FF, both large, while for $\pi^-$ the relevant terms always involve a suppressed factor. These asymmetries are larger than in the $pp$ case, as gluons play a negligible role here. Sizable signals are also found for the $\jt$ dependence of the asymmetry (see Fig.~5 of Ref.~\cite{DAlesio:2026pca}). Future measurements at the EIC would allow for testing in more depth the Collins FF universality and the TMD factorization for hadron-in-jet production processes.

\section{Conclusions}
\label{sec:conclusions}
We have studied the Collins azimuthal asymmetries for pion-in-jet production in polarized $pp$ and $ep$ collisions. We have shown that, within a simplified TMD approach and by adopting a recent extraction of the transversity and Collins function from SIDIS and $e^+e^-$ data, we provide a good description of the STAR data. This provides support for the universality of the Collins function, and shows that $pp$ data can improve and extend our knowledge of TMDs at large $x$.

By extending the analysis to the complementary $\ell p$ process, we have then obtained predictions for the EIC kinematics, going beyond LO through the inclusion of the quasireal photon exchange. Although this contribution is sizable, it does not spoil the dominance of quark-initiated channels, and asymmetries are predicted to be as large as $8-10\%$ within uncertainties and depending on the kinematics. Pion-in-jet production at the EIC will allow for a clearer access to the transversity distribution, including its still poorly known sea-quark component, and for a further test of TMD factorization and of the universality of the Collins function. 

Future work will address more refined parameterizations, including also TMD evolution, and a more accurate treatment of the jet fragmentation mechanism beyond the LO accuracy.

\section*{Acknowledgments}
%\vspace*{-2mm}\small
C.F.\ is supported by the European Union's Horizon Europe research and innovation programme under the Marie Sk\l{}odowska-Curie grant agreement n.~101150792 (STAT-TMDs). The work of U.D.\ is partially supported by Fondazione di Sardegna under the project ``Journey to the center of the proton'', No.~F23C25000150007 (University of Cagliari).  The work of M.Z.\ was supported by the U.S. Department of Energy, Office of Science, Office of Nuclear Physics under Contract No.~89243126CSC000213.
%The work of M.Z.\ was partially supported by the U.S.\ Department of Energy contract No.~DE-AC05-06OR23177, under which Jefferson Science Associates, LLC operates Jefferson Lab.

%\normalsize
\bibliographystyle{JHEP}
\setlength{\bibsep}{\itemsep}
\bibliography{references}

@article{Kang:2017btw,
    author = "Kang, Zhong-Bo and Prokudin, Alexei and Ringer, Felix and Yuan, Feng",
    title = "{Collins azimuthal asymmetries of hadron production inside jets}",
    eprint = "1707.00913",
    archivePrefix = "arXiv",
    primaryClass = "hep-ph",
    reportNumber = "JLAB-THY-17-2497",
    doi = "10.1016/j.physletb.2017.10.031",
    journal = "Phys.~Lett.~B",
    volume = "774",
    pages = "635",
    year = "2017"
}

@article{DAlesio:2020vtw,
    author = "D'Alesio, Umberto and Flore, Carlo and Prokudin, Alexei",
    title = "{Role of the Soffer bound in determination of transversity and the tensor charge}",
    eprint = "2001.01573",
    archivePrefix = "arXiv",
    primaryClass = "hep-ph",
    reportNumber = "JLAB-THY-19-3130",
    doi = "10.1016/j.physletb.2020.135347",
    journal = "Phys.~Lett.~B",
    volume = "803",
    pages = "135347",
    year = "2020"
}

@article{Collins:1992kk,
    author = "Collins, John C.",
    title = "{Fragmentation of transversely polarized quarks probed in transverse momentum distributions}",
    eprint = "hep-ph/9208213",
    archivePrefix = "arXiv",
    reportNumber = "PSU-TH-102",
    doi = "10.1016/0550-3213(93)90262-N",
    journal = "Nucl.~Phys.~B",
    volume = "396",
    pages = "161",
    year = "1993"
}

@article{Boglione:2024dal,
    author = "Boglione, Mariaelena and D'Alesio, Umberto and Flore, Carlo and Gonzalez-Hernandez, Jos\`e Osvaldo and Murgia, Francesco and Prokudin, Alexei",
    title = "{Simultaneous reweighting of Transverse Momentum Dependent distributions}",
    eprint = "2402.12322",
    archivePrefix = "arXiv",
    primaryClass = "hep-ph",
    reportNumber = "JLAB-THY-24-3995",
    doi = "10.1016/j.physletb.2024.138712",
    journal = "Phys. Lett. B",
    volume = "854",
    pages = "138712",
    year = "2024"
}

@article{Collins:1981uk,
    author = "Collins, John C. and Soper, Davison E.",
    title = "{Back-To-Back Jets in QCD}",
    reportNumber = "OITS-155",
    doi = "10.1016/0550-3213(81)90339-4",
    journal = "Nucl. Phys. B",
    volume = "193",
    pages = "381",
    year = "1981",
    note = "[Erratum: Nucl.Phys.B 213, 545 (1983)]"
}

@article{Collins:1984kg,
    author = "Collins, John C. and Soper, Davison E. and Sterman, George F.",
    title = "{Transverse Momentum Distribution in Drell-Yan Pair and W and Z Boson Production}",
    reportNumber = "CERN-TH-3923",
    doi = "10.1016/0550-3213(85)90479-1",
    journal = "Nucl. Phys. B",
    volume = "250",
    pages = "199--224",
    year = "1985"
}

@article{Metz:2002iz,
    author = "Metz, A.",
    title = "{Gluon-exchange in spin-dependent fragmentation}",
    eprint = "hep-ph/0209054",
    archivePrefix = "arXiv",
    doi = "10.1016/S0370-2693(02)02899-X",
    journal = "Phys. Lett. B",
    volume = "549",
    pages = "139--145",
    year = "2002"
}

@article{Collins:2004nx,
    author = "Collins, John C. and Metz, Andreas",
    title = "{Universality of soft and collinear factors in hard-scattering factorization}",
    eprint = "hep-ph/0408249",
    archivePrefix = "arXiv",
    doi = "10.1103/PhysRevLett.93.252001",
    journal = "Phys. Rev. Lett.",
    volume = "93",
    pages = "252001",
    year = "2004"
}

@article{Yuan:2007nd,
    author = "Yuan, Feng",
    title = "{Azimuthal asymmetric distribution of hadrons inside a jet at hadron collider}",
    eprint = "0709.3272",
    archivePrefix = "arXiv",
    primaryClass = "hep-ph",
    reportNumber = "RBRC-689",
    doi = "10.1103/PhysRevLett.100.032003",
    journal = "Phys. Rev. Lett.",
    volume = "100",
    pages = "032003",
    year = "2008"
}

@article{DAlesio:2010sag,
    author = "D'Alesio, Umberto and Murgia, Francesco and Pisano, Cristian",
    title = "{Azimuthal asymmetries for hadron distributions inside a jet in hadronic collisions}",
    eprint = "1011.2692",
    archivePrefix = "arXiv",
    primaryClass = "hep-ph",
    doi = "10.1103/PhysRevD.83.034021",
    journal = "Phys. Rev. D",
    volume = "83",
    pages = "034021",
    year = "2011"
}

@article{Anselmino:2007fs,
    author = "Anselmino, M. and Boglione, M. and D'Alesio, U. and Kotzinian, A. and Murgia, F. and Prokudin, A. and T{\"u}rk, C.",
    title = "{Transversity and Collins functions from SIDIS and $e^+ e^-$ data}",
    eprint = "hep-ph/0701006",
    archivePrefix = "arXiv",
    doi = "10.1103/PhysRevD.75.054032",
    journal = "Phys. Rev. D",
    volume = "75",
    pages = "054032",
    year = "2007"
}

@article{Bailey:2020ooq,
    author = "Bailey, S. and Cridge, T. and Harland-Lang, L. A. and Martin, A. D. and Thorne, R. S.",
    title = "{Parton distributions from LHC, HERA, Tevatron and fixed target data: MSHT20 PDFs}",
    eprint = "2012.04684",
    archivePrefix = "arXiv",
    primaryClass = "hep-ph",
    reportNumber = "IPPP/20/58",
    doi = "10.1140/epjc/s10052-021-09057-0",
    journal = "Eur. Phys. J. C",
    volume = "81",
    number = "4",
    pages = "341",
    year = "2021"
}

@article{deFlorian:2014xna,
    author = "de Florian, Daniel and Sassot, R. and Epele, Manuel and Hern\'andez-Pinto, Roger J. and Stratmann, Marco",
    title = "{Parton-to-Pion Fragmentation Reloaded}",
    eprint = "1410.6027",
    archivePrefix = "arXiv",
    primaryClass = "hep-ph",
    doi = "10.1103/PhysRevD.91.014035",
    journal = "Phys. Rev. D",
    volume = "91",
    number = "1",
    pages = "014035",
    year = "2015"
}

@article{STAR:2022hqg,
    author = "Abdallah, Mohamed and others",
    collaboration = "STAR",
    title = "{Azimuthal transverse single-spin asymmetries of inclusive jets and identified hadrons within jets from polarized $pp$ collisions at $\sqrt{s}$ = 200 GeV}",
    eprint = "2205.11800",
    archivePrefix = "arXiv",
    primaryClass = "hep-ex",
    doi = "10.1103/PhysRevD.106.072010",
    journal = "Phys. Rev. D",
    volume = "106",
    number = "7",
    pages = "072010",
    year = "2022"
}

@article{Salam:2008qg,
      author         = "Salam, G.~P. and Rojo, J.",
      title          = "{A Higher Order Perturbative Parton Evolution Toolkit (HOPPET)}",
      journal        = "Comput.~Phys.~Commun.",
      volume         = "180",
      year           = "2009",
      pages          = "120-156",
      doi            = "10.1016/j.cpc.2008.08.010",
      eprint         = "0804.3755",
      archivePrefix  = "arXiv",
      primaryClass   = "hep-ph",
      SLACcitation   = "%%CITATION = ARXIV:0804.3755;%%"
}

@article{STAR:2025xyp,
    author = "Aboona, Bassam and others",
    collaboration = "STAR",
    title = "{Energy Independence of the Collins Asymmetry in $p^\uparrow p$ Collisions}",
    eprint = "2507.16355",
    archivePrefix = "arXiv",
    primaryClass = "hep-ex",
    doi = "10.1103/jgsw-zhny",
    journal = "Phys. Rev. Lett.",
    volume = "135",
    number = "26",
    pages = "261902",
    year = "2025"
}

@article{DAlesio:2025jmr,
    author = "D'Alesio, Umberto and Flore, Carlo and Zaccheddu, Marco",
    title = "{Collins function for pion-in-jet production in polarized $pp$ collisions: a test of universality and factorization}",
    eprint = "2506.21959",
    archivePrefix = "arXiv",
    primaryClass = "hep-ph",
    reportNumber = "JLAB-THY-25-4388",
    doi = "10.1016/j.physletb.2025.140024",
    journal = "Phys. Lett. B",
    volume = "871",
    pages = "140024",
    year = "2025"
}

@article{Kniehl:1996we,
    author = "Kniehl, Bernd A. and Kramer, G. and Spira, M.",
    title = "{Large $p_T$ photoproduction of $D^{*\pm}$ mesons in $e p$ collisions}",
    eprint = "hep-ph/9610267",
    archivePrefix = "arXiv",
    reportNumber = "DESY-96-210, CERN-TH-96-274, MPI-PHT-96-103",
    doi = "10.1007/s002880050591",
    journal = "Z. Phys. C",
    volume = "76",
    pages = "689--700",
    year = "1997"
}

@article{vonWeizsacker:1934nji,
    author = "von Weizsacker, C. F.",
    title = "{Radiation emitted in collisions of very fast electrons}",
    doi = "10.1007/BF01333110",
    journal = "Z. Phys.",
    volume = "88",
    pages = "612--625",
    year = "1934"
}

@article{Williams:1934ad,
    author = "Williams, E. J.",
    title = "{Nature of the high-energy particles of penetrating radiation and status of ionization and radiation formulae}",
    doi = "10.1103/PhysRev.45.729",
    journal = "Phys. Rev.",
    volume = "45",
    pages = "729--730",
    year = "1934"
}

@article{DAlesio:2017nrd,
    author = "D'Alesio, Umberto and Flore, Carlo and Murgia, Francesco",
    title = "{Transverse single-spin asymmetries in $\ell \,p^\uparrow \to h \,X$ within a TMD approach: role of quasi-real photon exchange}",
    eprint = "1701.01148",
    archivePrefix = "arXiv",
    primaryClass = "hep-ph",
    doi = "10.1103/PhysRevD.95.094002",
    journal = "Phys. Rev. D",
    volume = "95",
    number = "9",
    pages = "094002",
    year = "2017"
}

@article{DAlesio:2026pca,
    author = "D'Alesio, Umberto and Flore, Carlo and Zaccheddu, Marco",
    title = "{Collins asymmetries for pion-in-jet production in polarized $\ell p$ collisions at the EIC}",
    eprint = "2605.02890",
    archivePrefix = "arXiv",
    primaryClass = "hep-ph",
    reportNumber = "JLAB-THY-26-4697",
    doi = "10.1016/j.physletb.2026.140543",
    journal = "Phys. Lett. B",
    volume = "878",
    pages = "140543",
    year = "2026"
}

\end{document}